\documentclass[useAMS,usenatbib]{mn2e}
\usepackage{aas_macros,graphicx,times,multirow}

\def \psr{PSR\, J1811$-$1736}

\title[Infrared observations of \psr]{Infrared observations of the candidate double neutron star system \psr}
\author[R. P. Mignani, et al.]
{\parbox{\textwidth}{R. P. Mignani$^{1,2}$\thanks{E-mail:rm2@mssl.ucl.ac.uk}, 
A. Corongiu$^{3}$, 
C. Pallanca$^{4}$, 
F. R. Ferraro$^{4}$} \\ \\
$^{1}$ Mullard Space Science Laboratory, University College London, Holmbury St. Mary, Dorking, Surrey, RH5 6NT, UK\\
$^{2}$ Kepler Institute of Astronomy, University of Zielona G\'ora, Lubuska 2, 65-265, Zielona G\'ora, Poland \\
$^{3}$ INAF Osservatorio Astronomico di Cagliari, localit\`a Poggio dei Pini, Strada 54, I--09012 Capoterra, Italy \\
$^{4}$ Dipartimento di Fisica e Astronomia, Universit\`a degli Studi di Bologna, via Ranzani 1, I--40127 Bologna, Italy \\
 }
\begin{document}

\date{Accepted 1988 December 15. Received 1988 December 14; in original form 1988 October 11}

\pagerange{\pageref{firstpage}--\pageref{lastpage}} \pubyear{2002}

\maketitle

\label{firstpage}

\begin{abstract}
PSR\,J1811$-$1736  ($P_{\rm s}$\,=\,104\,ms)  is  an old  ($\sim 1.89$ Gyrs)  binary pulsar  ($P_{\rm orb}$=18.8 d)  in a  highly eccentric orbit  ($e=\,0.828$)  with an  unidentified companion. Interestingly enough,  the pulsar  timing solution  yields an estimated companion mass $0.93\,M_{\odot}\le\,M_C\,\le\,1.5\,M_{\odot}$,  compatible  with that of a  neutron star.  As such, it is possible that PSR\, J1811$-$1736  is a double neutron  star (DNS) system, one  of the very  few discovered so far. This scenario can be investigated through deep optical/infrared (IR) observations. 
We used $J, H, K$-band images, obtained as part of the  UK Infrared Telescope (UKIRT) Infrared Deep Sky Survey (UKIDSS), and available in the recent Data Release 9 Plus, to search for its undetected companion of the PSR\,J1811$-$1736 binary pulsar. 
We detected a possible companion star to \psr\ within the $3 \sigma$ radio position uncertainty (1\farcs32), with magnitudes J=$18.61\pm 0.07$, H=$16.65\pm0.03$, and K$=15.46\pm 0.02$.  The star colours are consistent with either a main sequence (MS) star close to the turn-off or a lower red giant branch (RGB) star, at a pulsar distance of $\sim 5.5$  kpc and with a reddening  of $E(B-V)\approx 4.9$.  The star mass and radius would be compatible with the constraints on the masses and orbital inclination of the binary system inferred from the mass function and the lack of radio eclipses near superior conjunction. Thus,  it is possible that it is the companion to \psr. However, based on the star density in the field, we estimated a quite large chance coincidence probability of $\sim 0.27$  between the pulsar and the star, which makes the association unlikely.  No other star is detected within the $3\sigma$ pulsar radio position down to J$\sim 20.5$, H$\sim 19.4$ and K$\sim 18.6$, which would allow us to rule out a MS companion star earlier  than a mid-to-late M spectral type.
\end{abstract}


\section{Introduction}

The radio pulsar PSR\,J1811$-$1736 ($P_{\rm s}=104$ ms) was detected at 1374\,MHz (Lyne et al.\ 2000) during the Parkes Multibeam Pulsar survey (Manchester et al.\ 2001). It is in a binary system, with an orbital period $P_{\rm orb}$=18.8\,d and a high eccentricity $e$=0.828 (Corongiu et al.\ 2007). The updated timing parameters, including general relativistic effects, give a period derivative $\dot{P}_{\rm s}\,\sim\,9.01(5)\times10^{-19}$\,s\,s$^{-1}$ which yields a spin-down age $\tau_{\rm SD} \sim1.89\,\times 10^9$\,yr and a surface magnetic field  $B_{\rm surf}\sim9.8\times10^{9}$\,G. The  $P_{\rm s}$ and $\dot{P}_{\rm s}$  suggest that PSR\,J1811$-$1736 is a mildly-recycled pulsar, i.e. the spin-up phase via matter accretion from the companion  star was too short for the pulsar to reach a spin period of a few ms, typical of fully-recycled pulsars. A possible scenario is that the companion was an high-mass star which underwent a supernova explosion, itself turning into a neutron star (Bhattacharya \& van den Heuvel 1991). Thus, PSR\,J1811$-$1736 might be one of the $\approx$ 10 double neutron star  (DNS) systems out of the $\approx$ 2000 radio pulsars known to date (Manchester et al.\ 2005).
The DNS picture is reinforced by the limits on the companion mass, derived from the mass of the system $M_{\rm\,tot}\,=\,2.57\pm0.10\,M_{\odot}$ inferred from the measurement of the periastron advance $\dot{\omega}\,=\,0.0090^{\circ} \pm 0.0002^{\circ} $\,yr$^{-1}$ and from the mass function. For a pulsar mass $M_{\rm P}\,\ge\,1.17\,M_{\odot}$, larger than the minimum value inferred for a radio pulsar (PSR\, J1518+4904; Janssen et al.\ 2008), this yields a companion mass of $0.93\,M_{\odot}\le\,M_{\rm C}\,\le\,1.5 M_{\odot}$ (Corongiu et al.\ 2007), compatible with that of a neutron star.   

A conclusive piece of evidence that PSR\,J1811$-$1736  is a DNS would be the  detection of its  companion as  a radio pulsar, like in the double pulsar PSR\, J0737$-$3039A/B (Lyne et al.\ 2004). However, it escaped detection so far, perhaps because of an unfavourable beaming or because it is no longer in its active radio phase. Alternatively, a conclusive  piece of evidence would be the non-detection of the companion in deep optical/infrared (IR)  observations.  The pulsar companion is not detected  in the  Digitised Sky Survey (DSS)  down to $R \approx 22$ (Mignani 2000)  and in the 2 Micron All Sky Survey (2MASS) down to K$_{s} \approx15$, computed at the Lyne et al.\ (2000) and Corongiu et al.\ (2007) radio positions, respectively,  with the latter limit being quite uncertain owing to the much higher crowding in the pulsar field at IR wavelengths.   Such limits  would only  rule  out a giant companion but, for the allowed mass range they  would still be compatible with a mid to  late--type main sequence (MS) star, a white dwarf, or a neutron star.  No deep optical/near-IR  observations  of  PSR\, J1811$-$1736 have ever been performed so far. As suggested in Mignani (2000), given the substantial interstellar extinction towards the pulsar near-IR observations are more suited than the optical ones to set constraints on the companion star.

Here, we present the results of a new investigation of the \psr\ field using IR survey data much deeper than 2MASS. The observations and results are discussed in Sectn.\ 2, while the implications for the \psr\ companion are discussed in Sectn. 3.

\section{Infrared observations and results}

\begin{figure}
\centering
\includegraphics[height=8cm,bb=38 162 558 683,clip= ]{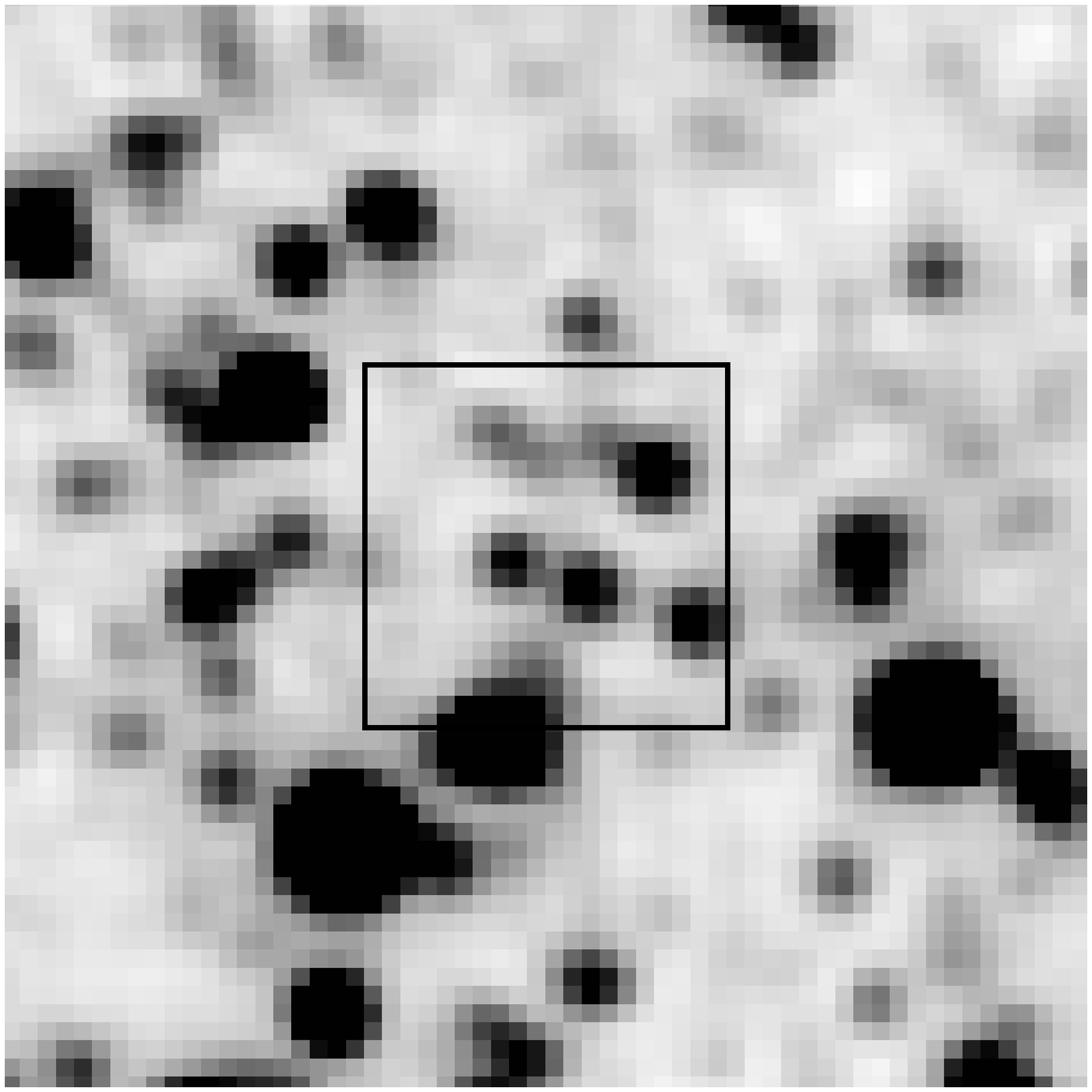}
\includegraphics[height=8cm, bb=37 162 559 681,clip=]{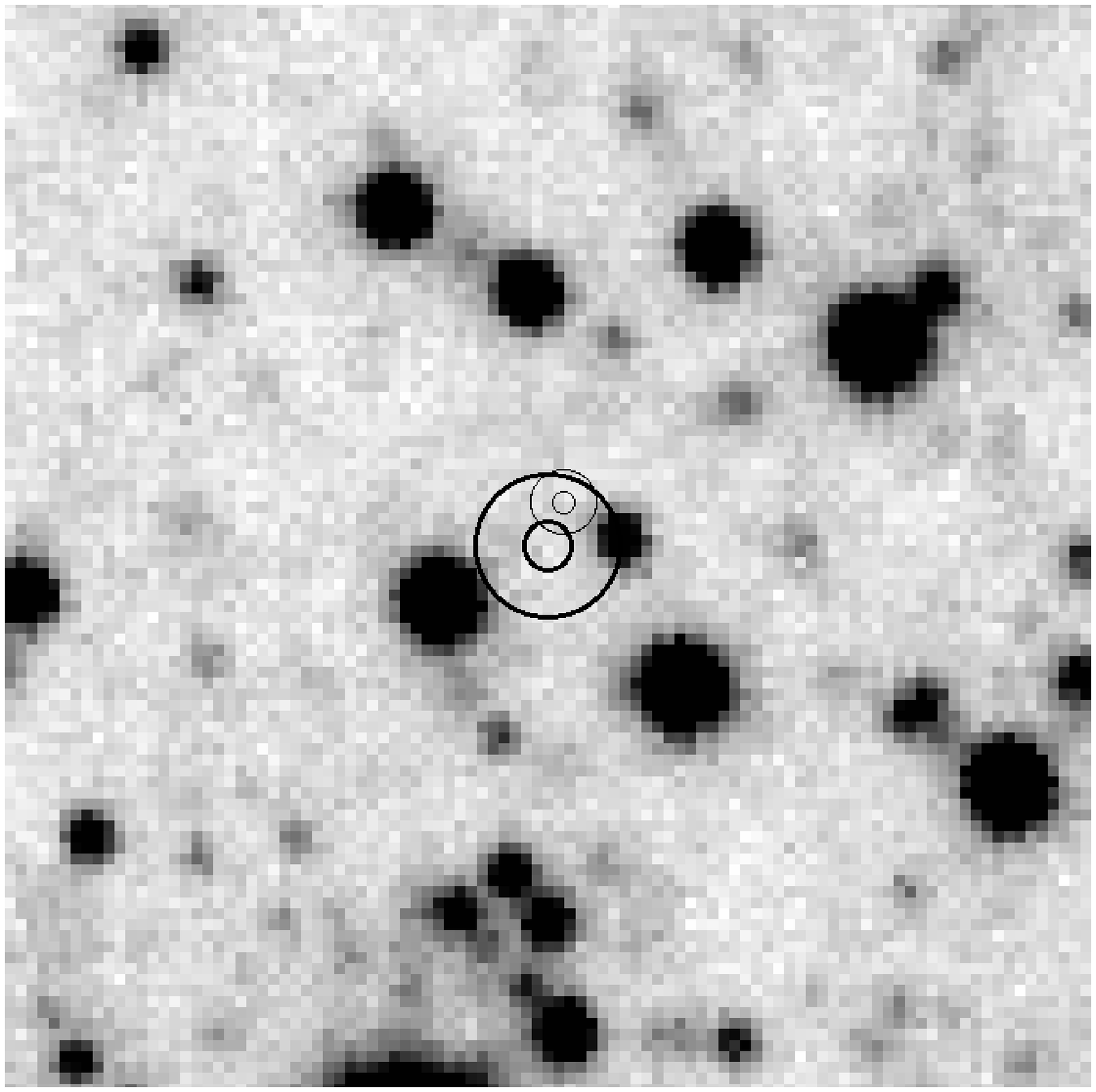}
  \caption{(\emph{top}) $1\arcmin \times 1\arcmin$ K$_s$-band image of the \psr\ field obtained from 2MASS. The $20\arcsec \times 20\arcsec$ square corresponds to the sky area shown in the bottom panel.  (\emph{bottom})  K$_s$-band image zoom of the same field obtained from UKIDSS.  North to the top, East to the left. The  radio positions of \psr\ computed with and without fitting its proper motion in the timing model (see Sectn.\ 2.2) are marked by the two sets of circles, drawn with thick and thin lines, respectively.  In both cases, the inner circles correspond to the $1\sigma$ (0\farcs44; 0\farcs19) uncertainty radii, while the outer circles correspond to the $3\sigma$ ($1\farcs32$; $0\farcs58$) ones.
  The star within the proper motion-corrected $3\sigma$ radio error circle (unresolved in the 2MASS image) has magnitudes J=$18.61\pm 0.07$, H=$16.65\pm0.03$, and K$_s=15.46\pm 0.02$. }
\end{figure}

\begin{figure*}
\centering
\includegraphics[width=16cm,clip=]{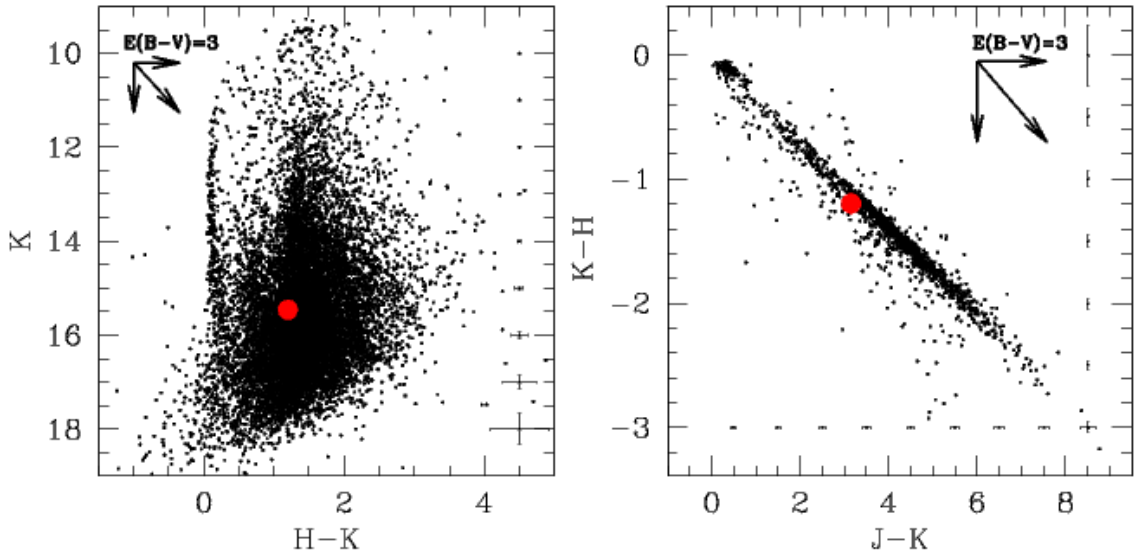}
  \caption{$K$--$(H-K)$ (\emph{left}) colour-magnitude diagrams of the \psr\ field obtained from the UKIDSS photometry of stars detected in a $10\arcmin \times 10\arcmin$ region around the \psr\ position. The star detected closest to the pulsar position (see Fig.\ 1) is marked in red. Average photometric errors in the pulsar field, for  different magnitude and colour bins, have are also plotted in the left and right panels respectively.  ($K-H$) vs. ($J-K$) colour-colour diagram of the same region.  The reddening vectors for an $E(B-V)=3$ are shown, as a reference. }
\end{figure*}

\subsection{Observation description}

No near-IR observations of the PSR\,J1811$-$1736 are available in either the ESO\footnote{http://archive.eso.org} or the Gemini\footnote{www.gemini.edu} Science Data Archives.  Thus, we searched for near-IR data of the \psr\ field in the image archive of the UK  Infrared Deep Sky Survey (UKIDSS) performed with the Wide Field Camera (WFCAM) at the  UK Infrared Telescope (UKIRT) at the Mauna Kea Observatory (Hawaii).   WFCAM (Casali et al.\ 2007) is a mosaic detector of four 2048$\times$2048 pixel Rockwell devices, with a pixel scale of 0\farcs4 and covering a field--of--view of 0.21 square degrees.  A general description of the UKIDSS survey is given in Lawrence et al.\ (2007). The UKIDSS survey  covers several regions, with a  different sky coverage,  and sensitivity limits in the $ZYJHK$ UKIRT photometric system (Hewett et al.\ 2006). The field of \psr\ is included in the Galactic Plane Survey (GPS;  Lucas et al.\ 2008) which covers about 1800 square degrees in $JHK$ down to  sensitivity limits which are
more than a factor of ten deeper than 2MASS. Like all the UKIDSS data, the GPS images are  processed through a dedicated pipeline (Hambly et al.\ 2008) developed and operated at Cambridge Astronomical Survey Unit (CASU) which performs basic reduction steps (dark subtraction, flat fielding),   image de-jittering, stacking, and mosaicing.  The pipeline also runs a source detection algorithm and produces  source catalogues. Astrometry  and photometry calibration are performed using 2MASS stars as a reference (Hodgkin et al.\ 2009). We searched for the reduced science images of the \psr\  field and associated object catalogues through the WFCAM Science Archive (WSA)\footnote{http://surveys.roe.ac.uk/wsa/} interface accessible via the Royal Observatory Edinburgh. We queried the most recent UKIDSS Data Release (version 9 Plus) made available on October 25th 2011.  The field was observed on July 18 2006. We downloaded $10\arcmin \times 10\arcmin$ $J$, $H$, $K$-band stacks around the pulsar position and the associated multi-band object catalogues.  

\begin{figure*}
\centering
\includegraphics[width=16cm,clip=]{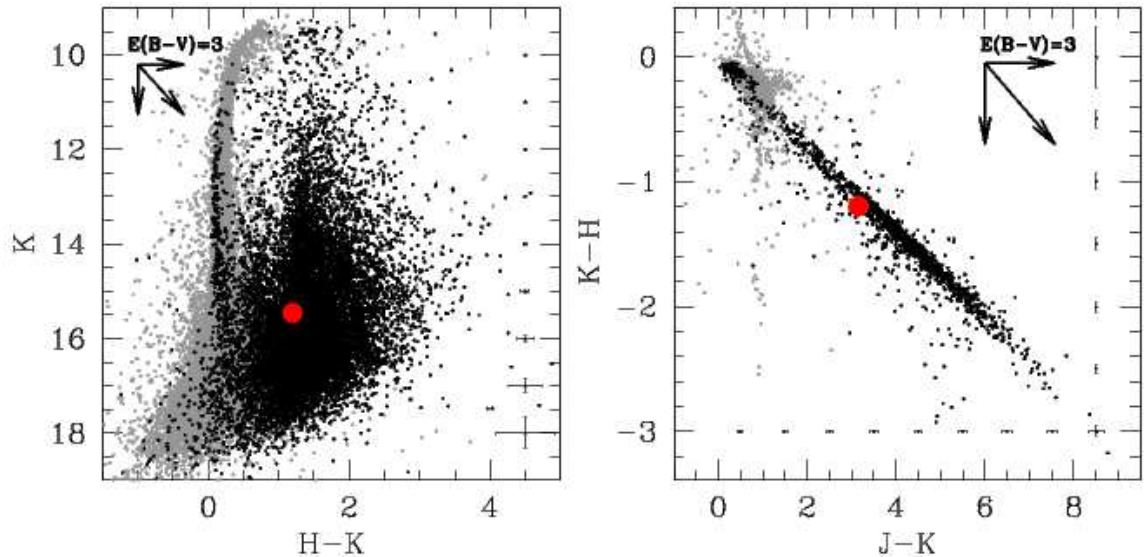}
  \caption{Same as Fig.\  2 but with UKIDSS data for the Baade's Window (light grey) overlaid for comparison.  }
\end{figure*}

\subsection{Pulsar astrometry}

For the search for the companion star to PSR\,J1811$-$1736, we assumed as a reference its radio timing coordinates. We note that the pulsar radio timing solution presented by Corongiu et al.\ (2007) is based on data taken in the epoch range MJD=50842--53624 and does not include the determination of the pulsar's proper motion, an essential parameter for recomputing its position at a given epoch. For this reason, we re-analysed the data presented in Corongiu et al.\ (2007) adding the proper motion to the timing model, to obtain a new radio timing position at a reference epoch (MJD=53624) closest to that of the UKIDSS observation (MJD= 53934). Thus, we obtained  $\alpha_{\rm J2000} =18^{\rm h}  11^{\rm m} 55\fs0385\pm 0\fs 0029$ and $\delta_{\rm J2000}  = -17^\circ 36\arcmin 38\farcs45\pm 0\farcs 41$ for the position and $\mu_{\alpha} cos(\delta) = 18.3 \pm 9.7$ mas\,yr$^{-1}$ and $\mu_{\delta}  = -176 \pm 100$ mas\,yr$^{-1}$ for the proper motion, where all quoted uncertainties are at $1 \sigma$ level. The extrapolated timing position at the epoch of the UKIDSS observation (MJD= 53934) is, then:  $\alpha_{\rm J2000} =18^{\rm h}  11^{\rm m} 55\fs054\pm 0\fs 009$ and $\delta_{\rm J2000}  = -17^\circ 36\arcmin 38\farcs60\pm 0\farcs 42$, with an uncertainty radius of 0\farcs44 (1$\sigma$) that accounts for the position  uncertainty due to the proper motion extrapolation.  
For comparison, by applying the same timing model as above but without adding the proper motion, we obtain $\alpha_{\rm J2000} =18^{\rm h}  11^{\rm m} 55\fs0337\pm 0\fs 0014$ and $\delta_{\rm J2000}  = -17^\circ 36\arcmin 37\farcs80\pm 0\farcs 19$, where the choice of the reference epoch (MJD=53624) within the range spanned by the timing data is, in this case, arbitrary. Although this position has a nominal uncertainty radius (0\farcs19; 1$\sigma$) that is smaller than obtained in the previous case, assuming it as a reference at the epoch of the UKIDSS observations would introduce an unknown systematic uncertainty due to the neglected pulsar proper motion.  For this reason,  it is formally less correct than the radio position obtained by fitting the proper motion, despite the latter having a larger uncertainty radius. Nonetheless, in the following section we conservatively consider both positions in our search for the  \psr\ counterpart.  In computing the overall uncertainty on the \psr\ position in the UKIDSS images we also accounted for systematics associated with the nominal accuracy on the UKIDSS astrometry calibration (0\farcs05 rms at low Galactic latitudes; Lawrence et al.\ 2007),  the internal astrometric accuracy of 2MASS ($\la$0\farcs2 for stars with $15.5 \le K \le 13$), and the accuracy on the link of 2MASS to the International Celestial Reference System (0\farcs015; Skrutskie et al.\ 2006).

\subsection{Results}

The UKIDSS $K$-band image of the \psr\ field is shown in Fig.\ 1 (bottom) compared to the corresponding 2MASS image (top).  For comparison, we plotted the two pulsar positions derived from the radio timing solution, with and without fitting the proper motion. As seen, no object is detected within the two $1\sigma$ radio position error circles (0\farcs44 and 0\farcs19 radii, respectively). However, a star ($K=15.46\pm 0.02$), unresolved in the 2MASS image but clearly detected in the much higher resolution UKIDSS one, is detected within the proper motion-corrected $3\sigma$ radio error circle (1\farcs32 radius).
Thus, its association with the pulsar cannot be ruled out a priori and needs to be investigated.  The star is also detected in the $J$ and $H$ bands, with  magnitudes of $J=18.61\pm 0.37$ and $H=16.65\pm0.03$.  No other star is detected at, or close to, the computed $3\sigma$ radio pulsar positions down to $3 \sigma$ limiting magnitudes of J$\sim 20.5$, H$\sim 19.4$ and K$\sim 18.6$, as computed from the rms of the sky background (Newberry 1991).  Given the high star density along the Galactic plane, however, the match can be the result of a chance coincidence. We computed this probability as $P=1-\exp(-\pi\rho r^2)$, where  $r$ ($\sim 1\farcs32$) is  the matching radius, assumed equal to the $3 \sigma$ uncertainty on the  proper motion-corrected pulsar radio position, and $\rho$ is  the  density of  stellar  objects within an area of $10\arcmin\times 10\arcmin$ around the pulsar.
We found that $\rho \sim 0.057$ arcsec$^{-2}$, which gives $P\sim 0.27$.  Such a high chance coincidence probability suggests that the star is likely unrelated to the pulsar, although we   need a direct piece of evidence to firmly rule out the association. 

\section{Discussion}

\subsection{The interstellar extinction in the pulsar direction}

We investigated whether the characteristics of the star detected close to the \psr\ position are compatible with it being its companion star. To this end, we tried to determine its spectral type from its colours. Fig.\ 2 shows the colour-magnitude  diagram (CMD) $(H-K)$--$K$ and the colour-colour diagram $(K-H)$ vs. $(J-K)$ built from the photometry of field stars detected within a $10\arcmin \times 10\arcmin$ area around the pulsar position, as derived from the UKIDSS object catalogue.  We also plotted the location of the star detected close to the \psr\ position (Fig.\ 1, bottom), whose location in both diagrams is consistent with the sequence of field stars. Thus, determining its spectral type from the comparison of its colours and flux with those of field stars is not obvious. Moreover, the determination of the star's intrinsic colours  is affected by the substantial interstellar extinction towards the pulsar, which is located in the Galactic plane ($l=12.828^{\circ}; b=0.435^{\circ}$).  In particular, the CMD  is very broadened, suggesting that the field is affected  by a quite high, and probably differential, extinction.

The interstellar extinction towards \psr\ is uncertain, and this affects our estimate of the upper limits on the companion star luminosity. A first estimate of the interstellar extinction can be derived from the integrated Hydrogen column density along the line of sight to the pulsar. This is $N_{\rm H}= (1.24-1.64) \times 10^{22}$,  as computed using the {\sc heasarc} tool {\tt webpimms}\footnote{http://heasarc.nasa.gov/cgi-bin/Tools/w3nh/w3nh.pl} according to the Dickey \& Lockman (1990) and  Kalberla et al.\ (2005) Hydrogen maps. This gives $E(B-V) = 2.2$--2.9, according to the relation of Predhel \& Schmitt (1995). However,  \psr\  is closer than the edge of the Galaxy, at a distance $D=5.7^{+0.84}_{-0.71}$\,kpc,  estimated from the radio pulse dispersion measure (DM=$476\pm5$ pc cm$^{-3}$; Corongiu et al.\ 2007)  and the Galactic free electron density along the line of sight  (Cordes \& Lazio 2002).  This would suggest a lower interstellar extinction. According to the Galactic extinction maps of Hakkila et al.\  (1997), the pulsar distance and Galactic coordinates would imply an interstellar extinction  $E(B-V)\,\sim\,1.9$.   
However, these estimates are only indicative mainly because of the uncertainties on the extinction maps on smaller angular scales. Unfortunately, the \psr\ field has not been observed in X-rays, so that no independent measurement of the interstellar extinction can be inferred from the hydrogen column density $N_{\rm H}$ directly derived from the fits to the X-ray spectra. In principle, an independent measurement of the $N_{\rm H}$ can be obtained from the DM itself assuming an average ionisation fraction of the interstellar medium (ISM) along the line of sight. In the case of \psr, a DM=$476\pm5$ pc cm$^{-3}$ would correspond to $N_{\rm H} \sim 1.5 \times 10^{22}$ cm$^{-2}$, for a 10\% ionisation fraction.  This  would imply an $E(B-V) \sim 2.7$. However, this method is usually applied to pulsars closer than $\sim 300$ pc (e.g., Pavlov et al. 2009; Tiengo et al.\  2011) and is  intrinsically affected by a much larger uncertainty for pulsars at larger distances, such as \psr.  

We tried to derive an independent estimate on the reddening  along the line of sight by comparing the CMDs and colour-colour diagrams of field stars with those in a reference region of very low reddening, such as the Baade's Window.   As we did for the \psr\ field, we extracted from the UKIDSS data the object catalogues relative to a $10\arcmin \times 10\arcmin$ area centred around the Baade's Window, for which we assumed the coordinates of the globular cluster NGC\, 6522: $\alpha_{\rm J2000} =18^{\rm h}  03^{\rm m} 34\fs08$ and  $\delta_{\rm J2000}  = -30^\circ 02\arcmin 02\farcs3$ (Di Criscienzo et al.\ 2006). Fig.\ 3, shows the same diagrams as in Fig.\ 2 but with the UKIDSS data for the BaadeÕs Window region overlaid. From the comparison of the two sets of diagrams, we derived an estimate of the interstellar extinction towards the pulsar. Firstly, we computed the average of the distribution in the colour-colour space for the Baade's Window region, applying a $3\sigma$ clipping.  Secondly, we did the same for the pulsar field but selecting a region of 50\arcsec\ radius around the pulsar position not to be affected by the differential extinction in the field. Then, from the difference between the two values we estimated an $E(B-V)=5.7\pm1.9$ for the pulsar field. This value is a factor of 2 larger than the Galactic interstellar extinction inferred from the  Hydrogen column density maps in the pulsar's direction. However, we note that the  $N_{\rm H}$ value reported above is a weighted average relative to a 1$^{\circ}$ radius area around the pulsar coordinates, which does not rule out the presence of patches of higher Hydrogen column density on angular scales smaller than 0.4$^{\circ}$, which are not resolved by the available maps. 

Indeed, it has been found that the interstellar extinction towards the Galactic bulge region is not uniform and shows strong variations, or granularities, on angular scales as small as 1\arcmin\ (see, e.g. Gosling et al.\ 2006). We used the UKIDSS J, H, and K$_s$-band images of the \psr\ field to measure the granularity of the  interstellar extinction in the region, following the method described in Gosling et al.\ (2006).  We considered a region of $500\arcsec \times 500\arcsec$ centred on the \psr\ position.
We divided the region in cells with dimension variable between 8\arcsec\ and 500\arcsec\ to sample the granularity in the field on different angular scales.  For each cell dimension, we calculated the $y$ parameter (see Eqn.\ 1 in Gosling et al.\ 2006),  which gives a quantitative estimate of the granularity of the field and is defined as the variance of the number of stars in all cell normalised to the mean number of stars per cell.  We computed the $y$ parameter for the J, H, and K$_s$-band images.  For the region considered in our analysis,  we found that the $y$ parameter, hence the granularity, decreases as a function of  wavelength (see Fig.\ 4), as in the case of one of the test fields used by Gosling et al.\ (2006). Thus, the high level of granularity in the pulsar field seems to be correlated with a high and variable reddening, explaining the large scatter in the CMD and colour-colour diagram of the field stars (Fig.\ 2). We note that the measured  angular scale of the granularity in the \psr\ field (see Fig.\  4) is comparable to the 50\arcsec\ radius region that we used to estimate the extinction in the pulsar's direction from the CMD and colour-colour diagram analysis (see previous paragraph).  Thus, we are confident that our procedure does not under/overestimate the assumed extinction value along the line of sight to the pulsar.

\begin{figure}
\centering
\includegraphics[width=8.7cm,bb=27 165 580 460,clip=]{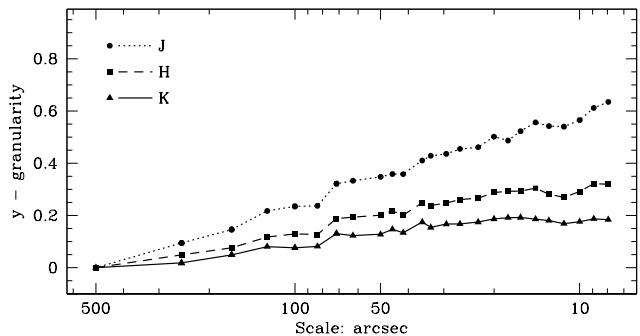}
  \caption{Measure of the granularity on the \psr\ field ($y$ parameter) computed using the method described in Gosling et al.\ (2006) as a function of the angular distance from the pulsar. The three curves correspond to the granularity measured in the J, H, and K$_s$-band images (see legenda). }
\end{figure}

\subsection{The candidate companion star}

Under the hypothesis that the star seen close to the \psr\ position is its companion, we tried to determine its spectral type assuming the range of reddening values computed above.  We considered a range of distances within the computed $1 \sigma$ uncertainty range on the \psr\ distance based on the DM ($D$=4.9--6.54 kpc), and an age range of 1--13.2 Gyr,  consistent with the ages of the stellar populations in the Galactic centre region (e.g. Zoccali et al.\ 2003), where the pulsar is located.  We also considered different values of the metallicity $Z$. For older populations, we considered both $Z=0.008$ and $Z=0.02$, while for the younger populations we considered only $Z=0.02$. Then, for different values of metallicity, age, and distance we determined the best $E(B-V)$ values in the range $E(B-V)=3.8$--7.6 that minimise the sum of the projected distances of the star location in the $J$--$(J-H)$ and $K$--$(H-K)$ CMDs from the isochrones computed from the stellar models of (Marigo et al.\ 2008; Girardi et al.\ 2010).  From the best combinations of metallicity, age, distance, and reddening,  we then derived the corresponding mass ($M_{\rm C}$) and radius ($R_{\rm C}$) of the candidate companion star to \psr, from the comparison with the isochrones. Finally,  we combined the computed mass of the candidate companion star  with the mass function of the pulsar system (Corongiu et al.\ 2007) and its total mass ($M_{\rm\,tot}\,=\,2.57\pm0.10\,M_{\odot}$).  In this way, we derived the pulsar mass ($M_{\rm P}$) and the orbital inclination angle $i$ of the system for each combination of metallicity, age, distance, reddening, and mass of the candidate companion star.  We filtered out combinations for which it resulted $\sin i>1$. We also selected the acceptable combinations for the conditions that the  companion mass  $0.93\,M_{\odot}\le\,M_{\rm C}\,\le\,1.5 M_{\odot}$  and the pulsar mass  $1.17\,M_\odot \le\,M_{\rm P}\,\le\, 1.6\,M_\odot$, as it results from the \psr\ timing analysis (Corongiu et al.\ 2007), where the lower limit  on  $M_{\rm P}$ corresponds to the minimum measured value for the mass of a neutron star (Janssen et al.\ 2008).  The different parameter combinations  are summarised in the first eight columns of Table\ 1 and 2.   These parameters are consistent,  for a reddening $E(B-V)=4.5$--5, with a companion star still on the main sequence (MS) and close to the turn-off or on the lower Red Giant Branch (RGB), with an inferred mass $M_{\rm C}\approx 1$--1.3 $M_{\odot}$ and radius $R_{\rm C} \approx 4.3$--5.8 $R_{\odot}$. 

\begin{table*}
\begin{center}
\caption{Best parameter combinations for the candidate companion star of \psr\ obtained from the comparison between its location in the CMDs and model isochrones. Columns list the star metallicity $Z$, age (in logarithmic units), distance ($D$), reddening, its mass  ($M_{\rm C}$) and radius ($R_{\rm C}$), the mass of the pulsar ($M_{\rm P}$) in solar units, and the system inclination angle ($i$), obtained from the estimated companion mass, the system mass function, and the total mass. The next two columns list the projected distance ($d$) of the pulsar to the centre of the companion on the plane perpendicular to the line of sight, normalised to the companion radius $R_{\rm C}$ for the two values of the orbital phase $\phi$ when the pulsar is observed closest to the superior conjunction $\phi\,=\,0.25$. The last column flags the acceptable configurations under the condition that $d/R_{\rm C}>1$ at both orbital phases (see text). }
\begin{tabular}{cccccc|cc|ccc} \hline\hline
$Z$ & Log(age) & D & $E(B-V)$     & $M_{\rm C}$  & $R_{\rm C}$ & $M_{\rm P}$ & $i$ &
\multicolumn{2}{c}{$d/R_{\rm C}$} & Flag   \\
&(yrs)        & (kpc) &            & ($M_{\odot}$) & ($R_{\odot}$) & ($M_{\odot}$) &
($\deg$)  & $\phi\,=\,0.191366$  & $\phi\,=\,0.277151$ & \\
\hline
0.020 & 10.12 & 6.54 & 4.70 & 0.97 & 5.78 & 1.60 & 76.11 & 0.65722 & 0.37243 & NO \\
0.020 & 10.12 & 5.70 & 4.72 & 0.97 & 5.02 & 1.60 & 76.17 & 0.75582 & 0.42761 & NO \\
0.020 & 10.12 & 4.99 & 4.76 & 0.97 & 4.41 & 1.60 & 76.28 & 0.85852 & 0.48427 & NO \\
0.020 & 10.10 & 6.54 & 4.70 & 0.99 & 5.77 & 1.58 & 73.56 & 0.69322 & 0.41833 & NO \\
0.020 & 10.10 & 5.70 & 4.73 & 0.99 & 5.03 & 1.58 & 73.63 & 0.79406 & 0.47843 & NO \\
0.020 & 10.10 & 4.99 & 4.76 & 0.99 & 4.41 & 1.58 & 73.72 & 0.90401 & 0.54356 & NO \\
0.020 & 10.08 & 6.54 & 4.70 & 1.00 & 5.77 & 1.57 & 71.36 & 0.72583 & 0.45845 & NO \\
0.020 & 10.08 & 5.70 & 4.73 & 1.00 & 5.03 & 1.57 & 71.42 & 0.83156 & 0.52463 & NO \\
0.020 & 10.08 & 4.99 & 4.77 & 1.00 & 4.41 & 1.57 & 71.53 & 0.94628 & 0.59574 & NO \\
0.020 & 10.06 & 6.54 & 4.70 & 1.01 & 5.76 & 1.56 & 69.47 & 0.75664 & 0.49420 & NO \\
0.020 & 10.06 & 5.70 & 4.74 & 1.01 & 5.04 & 1.56 & 69.47 & 0.86473 & 0.56480 & NO \\
0.020 & 10.06 & 4.99 & 4.77 & 1.01 & 4.40 & 1.56 & 69.62 & 0.98739 & 0.64330 & NO \\
0.020 & 10.04 & 6.54 & 4.71 & 1.02 & 5.78 & 1.55 & 67.74 & 0.78193 & 0.52452 & NO \\
0.020 & 10.04 & 5.70 & 4.74 & 1.02 & 5.03 & 1.55 & 67.74 & 0.89852 & 0.60273 & NO \\
0.020 & 10.04 & 4.99 & 4.78 & 1.02 & 4.41 & 1.55 & 67.88 & 1.02185 & 0.68407 & NO \\
0.020 & 10.02 & 6.54 & 4.71 & 1.03 & 5.76 & 1.54 & 66.11 & 0.81169 & 0.55665 & NO \\
0.020 & 10.02 & 5.70 & 4.75 & 1.03 & 5.03 & 1.54 & 66.30 & 0.92585 & 0.63339 & NO \\
0.020 & 10.02 & 4.99 & 4.78 & 1.03 & 4.41 & 1.54 & 66.30 & 1.05601 & 0.72244 & NO \\
0.020 & 10.00 & 6.54 & 4.72 & 1.05 & 5.78 & 1.52 & 64.72 & 0.83226 & 0.58041 & NO \\
0.020 & 10.00 & 5.70 & 4.75 & 1.05 & 5.03 & 1.52 & 64.72 & 0.95635 & 0.66695 & NO \\
0.020 & 10.00 & 4.99 & 4.79 & 1.05 & 4.42 & 1.52 & 64.72 & 1.08834 & 0.75900 & NO \\
0.008 & 10.00 & 6.54 & 4.83 & 0.97 & 5.77 & 1.60 & 76.04 & 0.65927 & 0.37429 & NO \\
0.008 & 10.00 & 5.70 & 4.85 & 0.97 & 5.02 & 1.60 & 76.14 & 0.75627 & 0.42821 & NO \\
0.008 & 10.00 & 4.99 & 4.88 & 0.97 & 4.40 & 1.60 & 76.28 & 0.86047 & 0.48537 & NO \\
  \hline \hline
\end{tabular}
	   \label{sim}
 \end{center}
\end{table*}

\begin{table*}
\begin{center}
\caption{Same as Tab. \ref{sim} but for ages smaller than 10 Gyrs. }
\begin{tabular}{cccccc|cc|ccc} \hline\hline
$Z$ & Log(age) & D & $E(B-V)$     & $M_{\rm C}$  & $R_{\rm C}$ & $M_{\rm P}$ & $i$ &
\multicolumn{2}{c}{$d/R_{\rm C}$} & Flag   \\
&(yrs)        & (kpc) &            & ($M_{\odot}$) & ($R_{\odot}$) & ($M_{\odot}$) &
($\deg$)  & $\phi\,=\,0.191366$  & $\phi\,=\,0.277151$ & \\
\hline
0.020 & 9.98 & 6.54 & 4.72 & 1.06 & 5.77 & 1.51 & 63.27 & 0.85842 & 0.60814 & NO \\
0.020 & 9.98 & 5.70 & 4.76 & 1.06 & 5.04 & 1.51 & 63.27 & 0.98276 & 0.69623 & NO \\
0.020 & 9.98 & 4.99 & 4.78 & 1.06 & 4.40 & 1.51 & 63.38 & 1.12323 & 0.79484 & NO \\
0.008 & 9.98 & 6.54 & 4.83 & 0.99 & 5.76 & 1.58 & 73.50 & 0.69529 & 0.42014 & NO \\
0.008 & 9.98 & 5.70 & 4.86 & 0.99 & 5.03 & 1.58 & 73.59 & 0.79471 & 0.47925 & NO \\
0.008 & 9.98 & 4.99 & 4.88 & 0.99 & 4.40 & 1.58 & 73.72 & 0.90607 & 0.54480 & NO \\
0.020 & 9.96 & 6.54 & 4.73 & 1.07 & 5.78 & 1.50 & 61.93 & 0.87993 & 0.63160 & NO \\
0.020 & 9.96 & 5.70 & 4.76 & 1.07 & 5.03 & 1.50 & 61.93 & 1.01114 & 0.72577 & NO \\
0.020 & 9.96 & 4.99 & 4.79 & 1.07 & 4.41 & 1.50 & 62.08 & 1.14991 & 0.82422 & NO \\
0.008 & 9.96 & 6.54 & 4.83 & 1.00 & 5.76 & 1.57 & 71.30 & 0.72801 & 0.46035 & NO \\
0.008 & 9.96 & 5.70 & 4.86 & 1.00 & 5.02 & 1.57 & 71.44 & 0.83287 & 0.52525 & NO \\
0.008 & 9.96 & 4.99 & 4.88 & 1.00 & 4.39 & 1.57 & 71.56 & 0.94999 & 0.59773 & NO \\
0.020 & 9.94 & 6.54 & 4.73 & 1.08 & 5.77 & 1.49 & 60.71 & 0.90254 & 0.65489 & NO \\
0.020 & 9.94 & 5.70 & 4.77 & 1.08 & 5.04 & 1.49 & 60.71 & 1.03326 & 0.74975 & NO \\
0.020 & 9.94 & 4.99 & 4.80 & 1.08 & 4.42 & 1.49 & 60.85 & 1.17504 & 0.85160 & NO \\
0.008 & 9.94 & 6.54 & 4.84 & 1.01 & 5.76 & 1.56 & 69.47 & 0.75664 & 0.49420 & NO \\
0.008 & 9.94 & 5.70 & 4.87 & 1.01 & 5.01 & 1.56 & 69.62 & 0.86716 & 0.56498 & NO \\
0.008 & 9.94 & 4.99 & 4.89 & 1.01 & 4.41 & 1.56 & 69.62 & 0.98515 & 0.64185 & NO \\
0.020 & 9.92 & 6.54 & 4.74 & 1.10 & 5.77 & 1.47 & 59.57 & 0.92229 & 0.67548 & NO \\
0.020 & 9.92 & 5.70 & 4.77 & 1.10 & 5.03 & 1.47 & 59.57 & 1.05797 & 0.77486 & NO \\
0.020 & 9.92 & 4.99 & 4.80 & 1.10 & 4.41 & 1.47 & 59.66 & 1.20467 & 0.88167 & NO \\
0.008 & 9.92 & 6.54 & 4.84 & 1.02 & 5.77 & 1.55 & 67.74 & 0.78329 & 0.52543 & NO \\
0.008 & 9.92 & 5.70 & 4.87 & 1.02 & 5.03 & 1.55 & 67.88 & 0.89589 & 0.59975 & NO \\
0.008 & 9.92 & 4.99 & 4.89 & 1.02 & 4.40 & 1.55 & 68.02 & 1.02117 & 0.68222 & NO \\
0.020 & 9.90 & 6.54 & 4.74 & 1.11 & 5.77 & 1.46 & 58.40 & 0.94257 & 0.69644 & NO \\
0.020 & 9.90 & 5.70 & 4.79 & 1.11 & 5.06 & 1.46 & 58.49 & 1.07305 & 0.79233 & NO \\
0.020 & 9.90 & 4.99 & 4.81 & 1.11 & 4.42 & 1.46 & 58.53 & 1.22751 & 0.90612 & NO \\
0.008 & 9.90 & 6.54 & 4.85 & 1.03 & 5.77 & 1.54 & 66.17 & 0.80928 & 0.55457 & NO \\
0.008 & 9.90 & 5.70 & 4.88 & 1.03 & 5.03 & 1.54 & 66.30 & 0.92585 & 0.63339 & NO \\
0.008 & 9.90 & 4.99 & 4.89 & 1.03 & 4.39 & 1.54 & 66.42 & 1.05819 & 0.72280 & NO \\
0.020 & 9.80 & 6.54 & 4.49 & 1.18 & 5.24 & 1.39 & 53.51 & 1.13065 & 0.86080 & NO \\
0.020 & 9.80 & 5.70 & 4.84 & 1.18 & 5.12 & 1.39 & 53.54 & 1.15658 & 0.88040 & NO \\
0.020 & 9.80 & 4.99 & 4.75 & 1.17 & 4.30 & 1.40 & 53.68 & 1.37393 & 1.04509 & YES \\
0.020 & 9.70 & 6.54 & 4.73 & 1.25 & 5.67 & 1.32 & 48.96 & 1.12253 & 0.87220 & NO \\
0.020 & 9.70 & 5.70 & 4.80 & 1.25 & 5.01 & 1.32 & 49.07 & 1.26832 & 0.98506 & NO \\
0.020 & 9.70 & 4.99 & 4.88 & 1.25 & 4.48 & 1.32 & 49.17 & 1.41625 & 1.09952 & YES \\
0.020 & 9.60 & 6.54 & 4.84 & 1.34 & 5.85 & 1.23 & 44.99 & 1.15094 & 0.90651 & NO \\
0.020 & 9.60 & 5.70 & 4.84 & 1.34 & 5.05 & 1.23 & 45.09 & 1.33147 & 1.04838 & YES \\
0.020 & 9.60 & 4.99 & 4.90 & 1.33 & 4.48 & 1.24 & 45.22 & 1.49824 & 1.17923 & YES \\
 \hline \hline
\end{tabular}
	   \label{sim2}
 \end{center}
\end{table*}

All the combination of parameters reported in Tab.\,\ref{sim} and \ref{sim2} have been further checked against the lack of eclipses at superior conjunction (orbital phase $\phi$ = 0.25) in the radio timing observations. This check is grounded on the fact that the pulsar cannot be eclipsed at a given orbital phase if the corresponding pulse's time of arrival (ToA) has been determined, since ToA determination strictly requires the detection of the pulse.  Hence, we calculated the orbital phases for each ToA presented in Corongiu et al.\  (2007), and we obtained that the closest available ToAs before and after superior conjunction correspond to an orbital phase $\phi$ = 0.191366 and $\phi$ = 0.277151, respectively. A parameters' combination is acceptable only if, at both orbital phases, the projected distance $d$ of the pulsar to the centre of the companion, computed on the plane perpendicular to the line of sight, is larger than the companion radius, i.e. $d/R_{\rm C}>1$.  Columns 9 and 10 of Tab.\,\ref{sim}  and \ref{sim2} report the values of  $d/R_{\rm C}$ for the two values of the orbital phase computed above, with the possible combinations flagged YES and NO for the cases $d/R_{\rm C}>1$ and $d/R_{\rm C}<1$, respectively. Our calculation shows that  for only  4 out of the 63 possible combinations of selected parameters the required condition  is satisfied at both orbital phases.  These combinations imply a $\approx 5$ Gyr companion star, with mass $M_{\rm C} \approx1.3M_{\odot}$ and radius  $R_{\rm C} \approx5R_{\odot}$, at a distance of $\sim 5.5$  kpc  and with a reddening $E(B-V)\approx 4.9$.  This corresponds to a pulsar mass $M_{\rm P} \approx1.3 M_{\odot}$ and inclination angle for the system of $\approx45^{\circ}$. Thus, according to the constraints on the masses and orbital inclination of the binary system, it is theoretically possible that the star detected at the radio position is, indeed, the companion to the pulsar. In this case, \psr\ would not be a DNS.  

\begin{table*}
\begin{center}
\caption{Name, timing ($P_{\rm s}$, $\dot{P}_{\rm s}$, $\tau_{\rm SD}$, $B_{\rm surf}$) and orbital parameters ($P_{\rm orb}$, $e$) for binary pulsars with identified companions and masses in the same range as the \psr\ companion.  The sample has been selected from the ATNF pulsar data base (Manchester et al.\ 2005).  Pulsar names are sorted in right ascension. The values of the companion masses $M_C$ are either directly measured or calculated from other parameters (e.g. post-keplerian parameters). The two extremes of the mass range for $M_C$  are computed assuming an inclination angle  $i=90^{\circ}$ and  $i=60^{\circ}$, respectively,  and a neutron star mass of 1.35 $M_\odot$.  $i=60^{\circ}$ is the median orbital inclination, for which there is an equal probability $P$ of having an inclination smaller or larger than $60^{\circ}$, i.e.  $P(i<60^{\circ})=P(i>60^{\circ})=0.5$. The last two columns indicate the companion type (WD, NS, MS) and whether the pulsar system is ordinary or recycled. }
\begin{tabular}{l|ccllrcccc} \hline\hline
Name & $P_{\rm s}$ & $\dot{P}_{\rm s}$ & $\tau_{\rm SD}$ & $B_{\rm surf}$ & $P_{\rm orb}$ & $e$ & $M_C$  & Companion & Ordinary (O)/ \\ 
           & (s)       & (s\,s$^{-1})$  &  (yrs) &  (G)          & (d)             &        & ($M_\odot$)  &              &  Recycled (R)\\  \hline
           J0514$-$4002A & 0.004991 & 1.17~$10^{-21}$ & 6.75~$10^{10}$ & 7.73~$10^{7}$ & 18.7852  & 8.880~$10^{-1}$ & 0.90--1.11 & WD & R \\
B0655+64 & 0.195671 & 6.85~$10^{-19}$ & 4.52~$10^{9}$ & 1.17~$10^{10}$ & 1.0287  & 7.500~$10^{-6}$ & 0.66--0.80 & WD & R \\
J0737$-$3039A & 0.022699 & 1.76~$10^{-18}$ & 2.04~$10^{8}$ & 6.40~$10^{9}$ & 0.1023  & 8.778~$10^{-2}$ & 1.24890  & NS & R \\
J1022+1001 & 0.016453 & 4.33~$10^{-20}$ & 6.01~$10^{9}$ & 8.55~$10^{8}$ & 7.8051  & 9.700~$10^{-5}$ & 1.05000  & WD & R \\
J1141$-$6545 & 0.393899 & 4.31~$10^{-15}$ & 1.45~$10^{6}$ & 1.32~$10^{12}$ & 0.1977  & 1.719~$10^{-1}$ & 1.02000  & WD & O \\
J1157$-$5112 & 0.043589 & 1.43~$10^{-19}$ & 4.83~$10^{9}$ & 2.53~$10^{9}$ & 3.5074  & 4.024~$10^{-4}$ & 1.18--1.46 & WD & R \\
J1337$-$6423 & 0.009423 & 1.95~$10^{-19}$ & 7.64~$10^{8}$ & 1.37~$10^{9}$ & 4.7853  & 2.004~$10^{-5}$ & 0.78--0.95 & WD & R \\
J1435$-$6100 & 0.009348 & 2.45~$10^{-20}$ & 6.05~$10^{9}$ & 4.84~$10^{8}$ & 1.3549  & 1.047~$10^{-5}$ & 0.88--1.08 & WD & R \\
J1439$-$5501 & 0.028635 & 1.42~$10^{-19}$ & 3.20~$10^{9}$ & 2.04~$10^{9}$ & 2.1179  & 4.985~$10^{-5}$ & 1.11--1.38 & WD & R \\
J1454$-$5846 & 0.045249 & 8.17~$10^{-19}$ & 8.78~$10^{8}$ & 6.15~$10^{9}$ & 12.4231  & 1.898~$10^{-3}$ & 0.86--1.05 & WD & R \\
J1518+4904 & 0.040935 & 2.72~$10^{-20}$ & 2.39~$10^{10}$ & 1.07~$10^{9}$ & 8.6340  & 2.495~$10^{-1}$ & 0.82--0.99 & NS & R \\
J1528$-$3146 & 0.060822 & 2.49~$10^{-19}$ & 3.87~$10^{9}$ & 3.94~$10^{9}$ & 3.1803  & 2.130~$10^{-4}$ & 0.94--1.15 & WD & R \\
B1534+12 & 0.037904 & 2.42~$10^{-18}$ & 2.48~$10^{8}$ & 9.70~$10^{9}$ & 0.4207  & 2.737~$10^{-1}$ & 1.35000   & NS & R \\
J1750$-$3703A & 0.111601 & 5.66~$10^{-18}$ & 3.12~$10^{8}$ & 2.54~$10^{10}$ & 17.3343  & 7.124~$10^{-1}$ & 0.58--0.69 & WD & R \\
J1756$-$2251 & 0.028462 & 1.02~$10^{-18}$ & 4.43~$10^{8}$ & 5.44~$10^{9}$ & 0.3196  & 1.806~$10^{-1}$ & 1.10--1.35 & NS & R \\
J1802$-$2124 & 0.012648 & 7.26~$10^{-20}$ & 2.76~$10^{9}$ & 9.69~$10^{8}$ & 0.6989  & 2.474~$10^{-6}$ & 0.78000  & WD & R \\
J1807$-$2459B & 0.004186 & 8.23~$10^{-20}$ & 8.06~$10^{8}$ & 5.94~$10^{8}$ & 9.9567  & 7.470~$10^{-1}$ & 1.20640   & WD & R \\
J1811$-$1736 & 0.104182 & 9.01~$10^{-19}$ & 1.83~$10^{9}$ & 9.80~$10^{9}$ & 18.7792 & 8.280~$10^{-1}$ & 0.85--1.04 & NS & R \\
B1820$-$11 & 0.279829 & 1.38~$10^{-15}$ & 3.22~$10^{6}$ & 6.29~$10^{11}$ & 357.7620  & 7.946~$10^{-1}$ & 0.65--0.78 & WD & O \\
J1829+2456 & 0.041010 & 5.25~$10^{-20}$ & 1.24~$10^{10}$ & 1.48~$10^{9}$ & 1.1760  & 1.391~$10^{-1}$ & 1.26--1.57 & NS & R \\
J1903+0327 & 0.002150 & 1.88~$10^{-20}$ & 1.81~$10^{9}$ & 2.04~$10^{8}$ & 95.1741  & 4.367~$10^{-1}$ & 1.03000   & MS & R \\
J1906+0746 & 0.144072 & 2.03~$10^{-14}$ & 1.13~$10^{5}$ & 1.73~$10^{12}$ & 0.1660 & 8.530~$10^{-2}$ & 0.80--0.98 & NS & O \\
B1913+16 & 0.059030 & 8.63~$10^{-18}$ & 1.08~$10^{8}$ & 2.28~$10^{10}$ & 0.3230 & 6.171~$10^{-1}$ & 1.3886$^{a}$   & NS & R \\
B2127+11C & 0.030529 & 4.99~$10^{-18}$ & 9.70~$10^{7}$ & 1.25~$10^{10}$ & 0.3353  & 6.814~$10^{-1}$ & 1.354$^{b}$   & NS & R \\
B2303+46 & 1.066371 & 5.69~$10^{-16}$ & 2.97~$10^{7}$ & 7.88~$10^{11}$ & 12.3395  & 6.584~$10^{-1}$ &1.16--1.43 & WD & O \\

   \hline \hline
\end{tabular}
	   \label{psrtab}
 \end{center}
 $^{a,b}$ The values of the companion mass is taken from Weisberg et al.\ (2010) and Jacoby et al.\ (2006), respectively, and are not yet implemented in the ANTF pulsar data base 
\end{table*}

A $\approx 5$ Gyr MS companion  would be compatible with the pulsar spin-down age ($\tau_{\rm SD} = 1.9$\,Gyr) but not much so with the recycling scenario and the pulsar orbital parameters. In principle, the high orbital eccentricity of \psr\ ($e=0.828$) can be seen as the signature of a supernova explosion that changed all binary system parameters. 
Since \psr\ is a recycled pulsar, the orbit must have been circularised during the recycling process (Battacharya \& van den Heuvel 1991) and its orbital eccentricity could have been produced by a second supernova explosion, i.e. that of the companion star.  In this case, both the spin period and the orbital eccentricity values of \psr\ would be the highest among DNSs and consistent with the correlation between these two parameters observed in such systems (Faulkner et al.\ 2005) and recovered under the hypothesis of a low-amplitude neutron star kick ($\sigma_v \sim 20$km\,s$^{-1}$) at birth (Dewi et al.\  2005). 

\begin{figure*}
\centering
\includegraphics[width=12.5cm,angle=270,clip=]{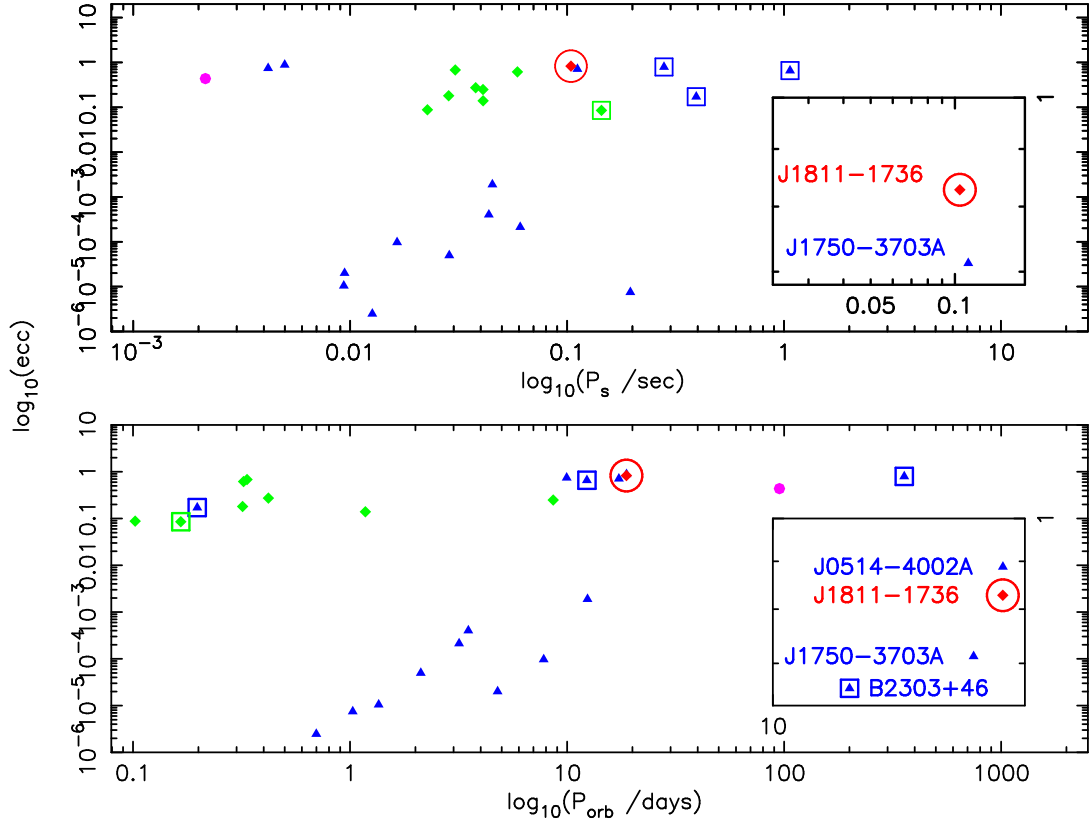}
  \caption{Orbital eccentricity $e$ vs. spin $P_{\rm s}$ (top panel) and orbital period  $P_{\rm orb}$ (lower panel) for the binary pulsars  in Table\ 3. Different symbols and colours correspond to different pulsar systems:  DNSs (green diamonds); pulsar--WD  (blue triangles); pulsar--MS (magenta filled circle). Squares highlight non--recycled pulsar systems. The location of \psr\ is marked by the circled red diamond. The inset shows a zoom of the plots around the location of \psr. Pulsar-systems falling closest to \psr\ in the $e$--$P_{\rm s}$ and $e$--$P_{\rm orb}$ planes are labeled and colour-coded.}
\end{figure*}

We investigated whether the MS companion scenario would be indeed compatible with the pulsar spin and orbital parameters. We compared in Fig.\ 5 the eccentricity, spin period, and orbital period of \psr\ to those of binary pulsars with identified companions, whose masses are in the same range as the \psr\ companion. We selected our sample from the Australia Telescope National Facility (ATNF) pulsar data base\footnote{http://www.atnf.csiro.au/people/pulsar/psrcat/} (Manchester et al.\ 2005).  Our sample is summarised in Table\ 3. In our analysis, we focused on the comparison with recycled binary pulsars only,  whose evolutionary path can be compared to that of \psr. As seen, the only known pulsar--MS star system in the selected companion mass range is PSR\, J1903+0327 (Khargharia et al.\ 2012). However, this is a fully--recycled ms-pulsar ($P_{\rm s}=2.5$ ms), whereas \psr\  is a mildly-recycled pulsars with a much longer spin period ($P_{\rm s}=104$ ms) and a much shorter orbital period  ($P_{\rm orb}=18.8$ d)  than  PSR\, J1903+0327 ($P_{\rm orb}=95.2$ d).  Moreover, the eccentricity of \psr\ ($e=0.828$) is much larger than  PSR\, J1903+0327 ($e=0.436$). Thus, there are no known pulsar--MS star systems in the selected companion mass range with spin and orbital parameters comparable to those of \psr.  This might suggest that such systems, if they do exists, are quite rare, although the very small sample currently available prevents us to draw firm conclusions.  There is one pulsar system, PSR\, J0514$-$4002A in the globular cluster NGC\,1851 (Freire et al.\ 2004), with a possible white dwarf (WD) companion (Freire et al.\ 2007), that has both orbital period and eccentricity comparable to  \psr\ (Fig.\ 5, lower panel).  However,  like PSR\, J1903+0327, also PSR\, J0514$-$4002A  is a fully--recycled pulsar, with a spin period $P_{\rm s}=4.99$ ms. Moreover,  since PSR\, J0514$-$4002A is in a globular cluster, its original companion might have been exchanged through a close encounter with another star in the cluster. Thus, the evolutionary history of this system might not be directly comparable to \psr.  A firm classification of the PSR\, J0514$-$4002A  companion would help to evaluate the pulsar--MS star scenario for \psr.

\subsection{Constraints on the nature of the companion star}

If the companion star of \psr\ is undetected in the UKIDSS data,  the derived upper limits on its flux can be used to constrain its nature.   From the interstellar extinction coefficients of Fitzpatrick (1999), an $E(B-V)=5.7\pm1.9$ would correspond to $A_J \sim 3.3$--6.6, $A_H \sim 2.0$--4.0 and $A_K \sim  1.4$--2.8. From our derived detection limits (J$\sim 20.5$, H$\sim 19.4$ and K$\sim 18.6$) these values imply extinction-corrected fluxes of $J_0 \ga 13.9$,  $H_0 \ga 15.4$, and  $K_0 \ga 15.8$, where we conservatively assumed the largest values of the interstellar extinction. At the estimated pulsar distance ($D=5.7^{+0.84}_{-0.71}$\,kpc) these values correspond to absolute magnitudes $M_J \ga 6.9$, $M_H\ga 8.4$, and $M_K\ga 8.8$, allowing us to rule out a companion of spectral type earlier  than a mid-to-late M-type MS star. Thus, our constraints on the companion star are far more compelling that those derived by Mignani (2000) on the basis of the DSS data alone. Moreover, those constraints should be now revised upward since the reddening towards the pulsar measured in this work is at least twice as large than assumed by Mignani (2000) from the Galactic extinction maps (Hakkila et al.\ 1997).  As a matter of fact, these were the only  resources available at the time to determine the reddening in the pulsar direction. 2MASS data of the pulsar field, which could be used to determine the reddening from the CMD technique, as we did with the UKIDSS data, were only released after the Mignani (2000) paper was published.  Our new limits are also not deep enough to rule out a WD companion star. As done in the previous section, we investigated whether the \psr\ spin and orbital parameters would fit those of recycled pulsar--WD systems. As seen from Fig.\ 5, most recycled pulsar--WD systems have circular orbits and both orbital and spin periods shorter than \psr.  There is only one recycled pulsar--WD system, PSR\, J1750$-$3703A in the globular cluster NGC\, 6397 (D'Amico et al.\ 2001),  that has both spin ($P_{\rm s}=0.111$ s) and orbital parameters ($P_{\rm orb}=17.33$ d; $e=0.712$) close to \psr.  Thus, it is possible that \psr\ is, indeed, a pulsar--WD system and not a DNS. However, since also PSR\, J1750$-$3703A is in a globular cluster, the same caveats as discussed above for  PSR\, J0514$-$4002A  apply in this case. The identification of other pulsar--WD systems  with spin and orbital parameter close to \psr, but located in the Galactic plane, would help to determine the nature of the PSR\, J1750$-$3703A system. We note that the other pulsar--WD system PSR\, B2303+46  (Dewey et al.\ 1985; Thorsett et al.\ 1993; van Kerkwijk \& Kulkarni 1999) with orbital parameters ($P_{\rm orb}=12.33$ d; $e=0.658$) similar to \psr\ (Fig.\ 5, lower panel) is not a recycled pulsar, which means that it is either on a different evolutionary path or evolutionary stage. 

\section{Summary and conclusions}

Using UKIDSS near-IR images, we detected a star, with magnitudes $J=18.61\pm 0.37$, $H=16.65\pm0.03$, and $K=15.46\pm 0.02$, within the $3\sigma$ radio position uncertainty of \psr.  In order to determine the star's spectral type, we estimated the reddening along the line of sight from the comparison of the CMDs of the stellar field to those of the Baade's Window, also built using UKIDSS data.  The reddening turns out to be at least twice as large as expected from the Galactic extinction maps.  At a pulsar distance of $\sim 5.5$  kpc, and for the estimated reddening  of $E(B-V)\approx 4.9$,  the star detected near to the radio position could be either a MS star close to the turn-off or a lower RGB star. The inferred mass ($\approx1.3M_{\odot}$) and radius ($\approx5 R_{\odot}$) of this star could be compatible with the pulsar mass function, the constraints on the pulsar and companion masses, and the lack of radio eclipses near superior conjunction. Thus, it is possible that this star is the pulsar companion, which would reject the DNS scenario for \psr.   if this is the case, this might be the first known example of a mildly--recycled pulsar--MS star system with companion mass in the $\sim0.9$--$1.5 M_{\odot}$ range, high eccentricity ($e>0.8$), and spin and orbital periods in the explored range.
However, we note that the computed chance coincidence probability of the candidate companion with the proper motion-corrected $3\sigma$ radio position is $P\sim 0.27$, which  suggests that it might, instead,  be an unrelated field star. 
A conclusive  piece of evidence to prove/disprove the association would come from IR spectroscopic observations of the candidate companion star along the orbital phase of the binary system and the comparison of its velocity curve with  that predicted by the orbital parameters of \psr.  Were the star confirmed to be its companion, this would drive new theoretical studies on the birth and evolution of neutron stars in binary systems, the formation of mildly-recycled radio pulsars, and the amplitude of neutron star kicks imparted by supernova explosions. 
On the other hand, were the star proved to be an unrelated field star, the identification of \psr\ as a DNS would remain an open issue.  
Our near-IR detection limits with UKIDSS only rule out a companion of spectral type earlier  than a mid-to-late M-type MS star. Deeper observations with 8m-class telescopes would enable us to push these limits down by about 4 magnitudes in each band. This would still not be enough to rule out any possible companion other than a neutron star, though, with a WD still compatible with the deepest achievable limits, unless the reddening is much lower than estimated in the current work. However, as shown in Sectn.\ 3.3, most recycled pulsar--WD systems in the Galactic plane do not fit  the  spin and orbital parameters of \psr.  
The detection of \psr\ in X-rays, like the other DNS PSR\, J1537+1155 (Durant et al.\ 2011), would be useful to independently constrain the reddening along the line of sight and put tighter constraints on the absolute luminosity of the companion. 
Finally, new radio observations of  \psr\ would be important to derive an updated radio-timing position and precisely measure the pulsar proper motion, for which we could only obtain a $2\sigma$ measurement using the current radio observation data base. A more precise radio-timing position of \psr\  will, then, enable us to revisit its association with its candidate companion star.

\section*{Acknowledgments}
We thank the anonymous members of the ESO Time Allocation Committee, whose suggestions triggered this work and the anonymous referee for his/her constructive comments to our manuscript.

\end{document}